\documentclass[a4paper,12pt]{article}
\usepackage{amsmath,amsfonts,amssymb}
\usepackage[latin1]{inputenc}
\usepackage[english]{babel}
\usepackage{graphicx}
\usepackage{rotating}
\usepackage{latexsym}
\usepackage{cite}
\usepackage{subfigure}

\newcommand{\be}{\begin{eqnarray}}
\newcommand{\en}{\end{eqnarray}}

\def\boxit#1{$\vcenter{\hrule\hbox{\vrule\kern3pt
     \vbox{\kern3pt\hbox{#1}\kern3pt}\kern3pt\vrule}\hrule}$}
\def\bigbox#1{$\vcenter{\hrule\hbox{\vrule\kern5pt
     \vbox{\kern5pt\hbox{#1}\kern5pt}\kern5pt\vrule}\hrule}$}
\def\Bigbox#1{$\vcenter{\hrule\hbox{\vrule\kern7pt
     \vbox{\kern7pt\hbox{#1}\kern7pt}\kern7pt\vrule}\hrule}$}

\begin{document}
\thispagestyle{empty}

\begin{flushright}
FTUAM 06/18\\
IFT-UAM/CSIC-06-56\\
\vspace*{10mm}
\end{flushright}
\begin{center}
{\Large \textbf{Symmetry breaking from Scherk-Schwarz compactification} }

\vspace{1.0cm}
M. Salvatori 

\vspace{1.0cm}
{\textit{Departamento de F\'{\i}sica Te\'{o}rica  and Instituto
 de F\'{\i}sica Te\'{o}rica , Universidad Aut\'{o}noma de Madrid,
 Cantoblanco, E-28049 Madrid, Spain.}}\\[0pt]
\end{center}

\vspace{2cm}
\begin{center}
{\bf Abstract}
\end{center}

We analyze the classical stable  configurations of an extra-dimensional gauge theory, in which the extra dimensions are compactified on a torus. Depending on the particular choice of gauge group and the number of extra dimensions, the classical vacua compatible  with four-dimensional Poincar\'e invariance and zero instanton number may have zero energy. For $SU(N)$ on a two-dimensional torus, we find and catalogue all possible degenerate zero-energy stable configurations in terms of continuous or discrete and continous parameters, for the case of trivial or non-trivial 't Hooft non-abelian flux, respectively. We then describe the residual symmetries of each vacua.

\clearpage

\section{Introduction}
\label{intro}

In the Standard Model, the explanation of the electroweak symmetry breaking pattern and the correlated hierarchy problem still remains an open issue. An intriguing extension is to consider a theory with extra space-like dimensions compactified on a non-simply connected manifold.  The compactification on non-simply connected manifolds, indeed, offers a new possibility of symmetry breaking: the Scherk-Schwarz (SS) mechanism \cite{ss}.  The non-local nature of this symmetry breaking protects the theory from  ultraviolet divergences  and makes it a promising candidate mechanism  to break the electroweak symmetry.

This idea has been widely investigated in the models of gauge-Higgs unifications \cite{early} in  five- \cite{5Drecenti} and six- \cite{6Drecenti} dimensional orbifold compactification, where the four dimensional scalars are identified with the internal components of a higher-dimensional gauge field. 

From the field theory point of view, a different and less explored possibility is to recover the idea of  gauge-Higgs unification in the context of {\it flux compactification}: compactification of the extra space-like dimensions on a manifold in which there exist a (gauge) background with a non-trivial field strength, compatible with SS periodicity conditions. 
Such type of compactification provides an alternative tool to obtain four-dimensional chirality \cite{Randjbar-Daemi:1982hi}. We investigate here some possible gauge symmetry breaking patterns that can be achieved in this context.

One simple example would be to consider a six-dimensional $U(N)$ gauge theory with the two extra dimensions  compactified on a two-dimensional torus $T^2$. As it is well known, the presence of a stable magnetic background associated with the abelian subgroup $U(1) \in U(N)$ and living only on the two extra dimensions, induces chirality. Furthermore, it affects the non-abelian subgroup $SU(N) \in U(N)$, giving rise to a non-trivial \textit{'t Hooft non-abelian flux} \cite{'tHooft:1979uj}. The latter can induce rich symmetry breaking patterns. While the case of trivial non-abelian 't Hooft flux is well-known in the literature \cite{Hosotani}, the field theory analysis and the phenomenology of the non-trivial non-abelian 't Hooft flux  has not been explored yet, except for a recent paper  \cite{ABGRS} analyzing mainly the case of $SU(2)$.

The main purpose of this paper is to find and classify all possible vacua 
and to describe the residual symmetries for an effective four-dimensional theory obtained from  a $SU(N)$ gauge theory on a six-dimensional space-time where the two extra dimensions are compactified on a torus, for both the cases of trivial and non-trivial 't  Hooft non-abelian flux. More in detail, in section \ref{section_T2vacuum} we provide a novel method to analyze the vacuum energy of a general  Lie group on an even-dimensional torus. For the case of $SU(N)$ on $T^2$, we re-obtain  a  well-known result \cite{Ambjorn:1980sm}: the stable vacua have always zero energy, including the case with coordinate-dependent periodicity conditions. In section  \ref{section_U}, we discuss  the relation between coordinate-dependent and constant transition functions and we find under which  conditions they are equivalent. For $SU(N)$, such result will allow to introduce in section \ref{section_back} what we denote as  background \textit{symmetric gauge}. In this gauge, we find and classify all the stable vacua and describe their symmetries for the case of trivial  't Hooft non-abelian flux as well as for the non-trivial case. 
Finally, in section \ref{conclusioni}, we conclude. 

\section{Vacuum energy of $SU(N)$ on $T^2$}
 \label{section_T2vacuum}

Consider a $SU(N)$ gauge theory on a $\mathcal{M}_4 \times T^2$ space-time.
In what follows, we will denote by $x$ the  coordinates of the four-dimensional Minkowski space $\mathcal{M}_4$ and by $y$ the extra space-like dimensions. 

A gauge field living on $T^2$  has to be periodic up to a gauge transformation under the fundamental shifts $\mathcal{T}_a : y \rightarrow y + l_a$ with $a=1,2$,  that define the torus\footnote{For simplicity, we consider an orthogonal torus, but all the results can be  generalized to a non orthogonal $T^2$.}:
\be
\mathbf{A_M} (x,y + l_a) &=& \Omega_a(y) \mathbf{A}_M (x,y) \Omega_a^\dagger(y) + \frac{i}{g} \Omega_a(y) \partial_M \Omega_a^\dagger(y) 
\label{boun_cond_A} \\
\mathbf{F}_{MN}(x,y+l_a) &=& \Omega_a(y) \mathbf{F}_{MN}(x,y) \Omega^\dagger(y) \,,
\label{boun_cond_F}
\en
where $M,N=0,1,...,5$, $a=1,2$ and $l_a$ is the length of the direction $a$. 
The eqs.(\ref{boun_cond_A})-(\ref{boun_cond_F}) are known as coordinate dependent Scherk-Schwarz compactification.
The transition functions $\Omega_a(y)$ are the embedding of the fundamental shifts in the gauge space and 
in order to preserve four-dimensional Poincar\'e invariance, they can only depend on the extra dimensions $y$.
Under a gauge transformation $S \in SU(N)$, the $\Omega_a(y)$ transform as
\be
\Omega_a' (y) = S(y+l_a) \, \Omega_a(y) \,S^\dagger(y) \,.
\label{omega_gauge}
\en
The transition functions  are constrained by the following consistency condition coming from the geometry:
\be
\Omega_1(y+l_2)\,\Omega_2(y)\,=\,e^{2 \pi i \frac{m}{N}}\,\Omega_2(y+l_1)\, \Omega_1(y) \,.
\label{cons_cond}
\en
The factor  exp$[2 \pi i m/N]$ is the representation  of the identity in the gauge space\footnote{A non-trivial value of $m$ is possible in the absence of field representations sensitive to the center of the group.}. The gauge invariant quantity $m=0,1,..,N-1$ (modulo $N$) is a topological quantity called \textit{non-abelian 't Hooft flux} \cite{'tHooft:1979uj}. In addition to the gauge transformations, the non-abelian 't Hooft flux is also invariant under the following group of transformations:
\be
\Omega_a (y) \,\rightarrow \,\Omega_a'(y) \,\equiv \,z_a \Omega_a (y) \,,
\label{omega_centro}
\en
where $z_a$ are elements of the center of $SU(N)$. This group is isomorphic to $\left(\mathbb{Z}_N\right)^2$.

The total Hamiltonian for a $SU(N)$ theory on a $\mathcal{M}_4 \times T^2$ space-time, reads
\be
H &=& \frac{1}{2} \,\,\int_{\mathcal{M}_4} d^4 x \, \int_{T^2} d^2 y \,\,\,\mathrm{Tr} \left[ \mathbf{F}^{MN} \mathbf{F}_{MN} \right] \nonumber \\
&=& \frac{1}{2} \,\,\int_{\mathcal{M}_4} d^4 x \, \int_{T^2} d^2 y\, \mathrm{Tr} \left[ \mathbf{F}^{\mu \nu} \mathbf{F}_{\mu \nu} + \mathbf{F}^{\mu a} \mathbf{F}_{\mu a} + \mathbf{F}^{ab} \mathbf{F}_{ab} \right] \,, 
\label{ener}
\en 
where here and in what follows, $\mu,\nu= 0,1,2,3$ and $a,b$ denote the extra coordinates. 
Since we are interested in configurations with $\mathbf{F}^{\mu \nu} \mathbf{F}_{\mu \nu}=0$ and which preserve four-dimensional Poincar\'e invariance (that is $\mathbf{F}_{\mu a }(x,y)=0$ and $\mathbf{F}_{ab}(x,y)= \mathbf{F}_{ab} (y)$), to minimize the expression in eq.~(\ref{ener}) reduces  to  minimize the quantity 
\be
H_{T^2} = \frac{1}{2} \,\,\int_{T^2} d^2 y \,\,\,\mathrm{Tr} \left[ \mathbf{F}^{ab} (y) \,\mathbf{F}_{ab} (y) \right] \geq 0 \,.
\label{enerT2}
\en
The latter inequality follows from the fact that 
we are working on an Euclidean manifold.

We will show that the vacuum energy is always zero, \textit{i.e.}  $\langle\mathbf{F}_{ab}\rangle=0$, including the case of coordinate-dependent periodicity conditions. This result reflects the non-existence of topological quantities for a $SU(N)
$ gauge theory on a $T^2$. 

Let us consider the issue for the more general case of a
Lie gauge group $\mathcal{G}$ on an even dimensional torus ($T^{2n}$ with the integer $n\geq1$), in order to pinpoint the dependence of the result on the choice of the gauge group and of the number of extra dimensions.

Parametrize  the $(4+2n)$-dimensional gauge field $\mathbf{A}_M$ as 
\be
\left\{
\begin{array}{lcl}
\mathbf{A}_{\mu}(x,y) &=&  A_\mu(x,y) \\
\mathbf{F}_{\mu \nu}(x,y) &=& F_{\mu \nu}(x,y)  
\end{array} 
\right. \,, \,\,\,
\left\{
\begin{array}{lcl}
\mathbf{A}_{a}(x,y) &=& B_a(y) + A_a(x,y)  \\
\mathbf{F}_{ab}(x,y) &=& G_{ab}(y) + F_{ab}(x,y) 
\end{array}
\right. \,,
\label{sosti}
\en
where the background $B_a(y)$ has the following properties:
\begin{itemize}
\item[i)] It is a solution of the $2n$ dimensional Yang-Mills equations of motion.
\item[ii)] It has non-trivial field strength.
\item[iii)] It is compatible with the periodicity conditions of eqs.(\ref{boun_cond_A})-(\ref{boun_cond_F}).
\end{itemize}
$A_\mu(x,y)$, $A_a(x,y)$ are the fluctuations fields. The background and fluctuation field strengths are defined as
\be
\label{field_st}
G_{ab} &=& \partial_a B_b -\partial_b B_a - i g \left[ B_a , B_b \right] \,, \nonumber \\ 
F_{\mu \nu} &=& \partial_\mu A_\nu - \partial_\nu A_\mu - i  g \left[ A_\mu , A_\nu \right] \,, \\
F_{ab} &=& D_a A_b - D_b A_a - i  g \left[ A_a , A_b \right] \,.\nonumber
\en
In eq.~(\ref{field_st}), $D_a A_b$ denotes the background covariant derivative  
\be
D_a A_b = \partial_a A_b - i  g \left[ B_a , A_b \right] \,,
\label{dev_cov}
\en
satisfying 
\be
\left[ D_a, \,D_b\right]\,=\,- i\, g \, G_{ab} \,.
\label{commu}
\en
Now $a,b=1,...,2n$.  For what follows, notice that for a non-simple gauge group, a solution of the classical Yang-Mills equations of motion can be  associated  to  generators belonging to the normal subgroup of the algebra. Such background $B_a$  satisfies $\left[B_a, A_b\right]=0$ and,  therefore, the covariant derivatives with respect to it reduce to ordinary derivatives.

Generalizing the discussion in ref.\cite{diag_field_strenght},  we diagonalize the background field strength with respect to the Lorentz indices. The first step is to perform an appropriate $O(2n)$ rotation able to write  the $2n \times 2n$ matrix  $G_{ab}(y)$ as\footnote{It follows from the fact that on an Euclidean flat space as $T^{2n}$, the  non-trivial coordinate dependence of $G_{ab}$  is completely determined  only by the gauge indices as it can be proved  using  the periodicity conditions of eqs.(\ref{boun_cond_A})-(\ref{boun_cond_F}) and the Yang-Mills equations of motion on a flat space.} 
\be
\begin{footnotesize}
G_{ab}  =  \left(
\begin{array}{cc}
0 & 
\begin{array}{cccc}
f_1 (y)  & 0 & ... & 0 \\
0 & f_2 (y) & ... & 0 \\
... & ... & ... & ... \\
0 & 0 & .. & f_n (y)
\end{array} \\
\begin{array}{cccc}
-f_1 (y) & 0 & ... & 0 \\
0 & -f_2 (y) & ... & 0 \\
... & ... & ... & ... \\
0 & 0 & .. & -f_n (y) 
\end{array} & 0
\end{array}
\right)
\end{footnotesize} \,,
\label{f_peq}
\en
where $f_i(y)$ for $i=1,...,n$ are matrices belonging to the adjoint representation of the gauge group $\mathcal{G}$.
The second step is to  introduce the  complex basis $\{z_i, \overline{z}_i \}$ defined as
\be
z_i = \frac{1}{\sqrt{2}} \left( y_i \,+ \,\,i \,y_{n+i} \right) \,,\hspace{1cm}
\overline{z}_i = \frac{1}{\sqrt{2}} \left( y_i \,- \,\,i \,y_{n+i} \right) \,,
\en
for $i = 1, ..., n$. In this basis, the background field strength is diagonal in the Lorentz space  
\be
G_{ab} = \mathrm{Diag} \left[ i f_1 (z), - i f_1(z), i f_2(z), -i f_2(z),..., i f_n(z), -i f_n (z)
\right] \,,
\en 
and  the commutators between the covariant derivatives in eq.~(\ref{commu}), reduce to 
\be
\left[ D_{z_i} , D_{z_j} \right] &=&\left[ D_{\overline{z}_i} , D_{\overline{z}_j} \right] =0 \,,\nonumber \\
\left[ D_{z_i} , D_{\overline{z}_j} \right] &=&  \,g\, f_i(z)  \,\delta_{ij} \,\,\,\,\,, \hspace{1cm} \forall \,i,j=1,...,n \,.
\label{comm_diag}
\en
We  introduce the following gauge fixing Hamiltonian compatible with the $2n$-dimensional generalization of the periodicity  conditions in eqs.(\ref{boun_cond_A})-(\ref{boun_cond_F})
\be 
H_{g.f.} = \int_{T^{2n}} \,d^nz\,d^n\overline{z}\,\,\mathrm{Tr} \left[\sum_{i=1}^n D_{z_i} A^{\overline{z}_i} + D_{\overline{z}_i} A^{z_i} \right]^2 \,.
\en
Denote $H_{T^{2n}}$ the $2n$-dimensional generalization of the Hamiltonian in eq.~(\ref{enerT2}).
Using eq.~(\ref{sosti}), the expansion of  $H_{T^{2n}} + H_{g.f.}$ up to second order in the perturbation fields $A_a(x,y)$ reads
\be
H_{T^{2n}} + H_{g.f.}= H^{(1A)} + H^{(2A)} + \mathcal{O}\left(A^3\right) \nonumber
\en
where
\be
\label{statio}
H^{(1A)} &=& - 2 \,\sum_{i=1}^n \,\, \int_{T^{2n}} \,d^nz\,d^{n}\overline{z} \,\,\mathrm{Tr} \left[ A^{z_i} D^{\overline{z}_i} G_{z_i \overline{z}_i} \,+\, A^{\overline{z}_i} D^{z_i} G_{\overline{z}_i z_i} \right] \,\,, \\
H^{(2A)} &=&  \, \sum_{i=1}^n \,\, \int_{T^{2n}} \,d^{n}z \,d^n\overline{z}\,\,\mathrm{Tr} \left[ A^{z_i} \mathcal{M}^2_{\overline{z}_i z_i} A^{\overline{z}_i} + A^{\overline{z}_i} \mathcal{M}^2_{z_i \overline{z}_i } A^{z_i} \right] \,.
\label{second}
\en
The operators $\mathcal{M}^2_{z_i\overline{z}_i}$ and $\mathcal{M}^2_{\overline{z}_i z_i}$ in eq.~(\ref{second}) are given by
\be
\label{massa_1}
\mathcal{M}^2_{\overline{z}_i z_i} &\equiv& \,\sum_{k=1}^n \Sigma_k \,+\,2\,\Gamma_i  \,\,,\\
\mathcal{M}^2_{z_i \overline{z}_i} &\equiv& \,\sum_{k=1}^n \Sigma_k\,-\,2\,\Gamma_i \,, 
\label{massa_2}
\en
where
\be
\label{Sigma}
\Sigma_{i} &\equiv& -\, \left\{ D_{z_i},\, D_{\overline{z}_i} \right\} \,\,,  \\
\Gamma_{i} &\equiv&  \,\left[ D_{z_i},\,D_{\overline{z}_i}\right] \,\,,\hspace*{2cm} \forall i=1,...,n\,. 
\label{Gamma}
\en
The background $B_a$ is then seen to be stable if and only if it is stationary, \textit{i.e.}
$H^{(1A)}=0$, and the eigenvalues of the operators defined in  eqs.(\ref{massa_1})-(\ref{massa_2}) are all  semi-positive. \\

Since $B_a$ is a solution of the classical equations of motion, it is stationary  by construction. 

In order to discuss the sign of the eigenvalues of the operators in eqs.(\ref{massa_1})-(\ref{massa_2}),  we recall that
\begin{itemize}
\item $\forall i=1,..,n$, the operators $\Sigma_i$ are defined semi-positive:
\be
\label{positive}
\hspace*{-0.4cm}\Sigma_i = - D_{z_i} D_{\overline{z}_i} - D_{\overline{z}_i} D_{z_i}&=& | D_{z_i} |^2 + | D_{\overline{z}_i} |^2 \geq 0 \,,
\en 
since $\left( D_{z_i} \right)^\dagger = - D_{\overline{z}_i}$ and $\left( D_{\overline{z}_i} \right)^\dagger = - D_{z_i}$.
\item The background $B_a$ satisfies the Yang-Mills equations of motion and then  the operators  $ \Sigma_i$, $ \Gamma_i$ commute. Consequently, there exists a basis that diagonalizes  simultaneously (with respect to the gauge indices) these operators.
We denote 
with $\mathbf{|}\,\lambda_{\Sigma_i},\lambda_{\Gamma_i} \,\mathbf{\rangle}$ the elements of such basis
satisfying
\be
\hspace{-0.8cm} \Sigma_k \,\,\,\, \vert \lambda_{\Sigma_i}, \lambda_{\Gamma_i}\,\rangle &=& \,\lambda_{\Sigma_k} \,\,\,| \lambda_{\Sigma_i},\lambda_{\Gamma_i}\,\rangle \,\,, \nonumber\\
\hspace{-0.8cm} \Gamma_k \,\,\,\, \vert \lambda_{\Sigma_i}, \lambda_{\Gamma_i}\,\rangle &=& \,\lambda_{\Gamma_k} \,\,\,| \lambda_{\Sigma_i},\lambda_{\Gamma_i}\,\rangle  \,\,,\nonumber
\en
for any $k=1,..,n$. 
\end{itemize}
We start analyzing the eigenvalues of the operators of eqs.(\ref{massa_1})-(\ref{massa_2}) associated to the elements $|\lambda_{\Sigma_i},\lambda_{\Gamma_i}\,\rangle$ belonging to the  subspace characterized by $\lambda_{\Gamma_i} = 0$ $\forall i=1,..,n$, that is to the subspace in which $\left[D_{z_i} , D_{\overline{z}_i} \right] = 0$, $\forall i=1,..,n$.  All the elements of this subspace have semi-positive defined eigenvalues since  eqs.(\ref{massa_1})-(\ref{massa_2}) reduce to
\be
\mathcal{M}^2_{\overline{z}_i z_i} = \mathcal{M}^2_{z_i \overline{z}_i} &\equiv& \sum_{k=1}^n \Sigma_k \geq 0 \,. 
\label{eig_semp}
\en
Notice that for the case of a non simple gauge group with background such that $\left[B_a, A_b\right]=0$, the subspace $\lambda_{\Gamma_i}=0$
coincides with the whole space. 

Consider, now,  the subspace associated to eigenvalues  $\lambda_{\Gamma_i} \neq0$.
It can be analyzed using the analogy with the harmonic oscillator, \textit{i.e.} using the non-trivial commutation rules in eq.~(\ref{Gamma}). 
The vacuum $|0\rangle$ is characterized by 
\be
&- D_{\overline{z}_i} D_{z_i}|0\rangle &= 0 \hspace{0.8cm} \mathrm{if} \,\,\,\lambda_{\Gamma_i} < 0  \nonumber \\
&- D_{z_i} D_{\overline{z}_i}|0\rangle &= 0 \hspace{0.8cm} \mathrm{if} \,\,\,\lambda_{\Gamma_i} > 0 \\
&- D_{z_i} D_{\overline{z}_i}|0\rangle =- D_{\overline{z}_i} D_{z_i}|0\rangle &= 0 \hspace{0.8cm} \mathrm{if} \,\,\,\lambda_{\Gamma_i} =0 \,. \nonumber
\en
For simplicity, we will discuss explicitly the subspace associated to the elements for which all $\lambda_{\Gamma_i} \neq 0$ are positive\footnote{The subspaces associated to eigenstates for which some $\lambda_{\Gamma_i} < 0$  can be obtained from the following reasoning interchanging $z_i \leftrightarrow  \overline{z}_i$ for those indices $i$ such that $\lambda_{\Gamma_i} < 0$.}.
The vacuum is, therefore,  defined 
\be
-D_{z_i} D_{\overline{z}_i}  |0\rangle = 0 \,,
\label{00}
\en
for all $i$ associated to  $\lambda_{\Gamma_i}\geq0$. Introduce the notation $\Sigma_i |0\rangle = \lambda_{\Sigma_i}^0 |0\rangle $ and $\Gamma_i |0\rangle = \lambda_{\Gamma_i}^0 |0\rangle$. Since $-D_{z_i} D_{\overline{z}_i} = 1/2 \left( \Sigma_i - \Gamma_i \right)$, eq.~(\ref{00}) implies 
\be
\lambda_{\Sigma_i}^0=\lambda_{\Gamma_i}^0 \,.
\label{import}
\en
The eigenvalues of the operators in eqs.(\ref{massa_1})-(\ref{massa_2}) associated to the vacuum $|0\rangle$ read 
\be
\mathcal{M}^2_{\overline{z}_i z_i}\,|0\rangle &=& \left(\sum_{k= 1}^n \lambda^0_{\Sigma_k} +\,2 \lambda^0_{\Gamma_i} \right) |0\rangle \,,
\label{massa_1_0}\\
\mathcal{M}^2_{z_i \overline{z}_i}\,|0\rangle &=& \left(\sum_{k= 1}^n \lambda^0_{\Sigma_k} -\,2 \lambda^0_{\Gamma_i} \right) |0\rangle \,.
\label{massa_2_0}
\en
Since $\lambda_{\Sigma_k}\geq 0$ for any $k=1,...,n$,
the right hand side (RHS) of eq.~(\ref{massa_1_0}) is always positive for $\lambda_{\Gamma_i} > 0$. On the contrary, the sign of the eigenvalue in the RHS of eq.~(\ref{massa_2_0}) is not determined \textit{a priori} for the general case of a Lie gauge group $\mathcal{G}$ on $T^{2n}$. 

Focusing on the case of $SU(N)$ on $T^2$, that is $n=1$, eq.~(\ref{massa_2_0}) reduces to
\be
\mathcal{M}^2_{z \overline{z}}\,|0\rangle\,=\, - \lambda_{\Gamma}^0\,|0\rangle \hspace*{1cm}\mathrm{with} \,\,\,\,\, \lambda_{\Gamma}^0  > 0 \,.
\label{massa_1_0_SU_T2}
\en
In this case, a background with a non-trivial field strength is, therefore, always unstable, since the operators defined in eqs.(\ref{massa_1})-(\ref{massa_2}) always admit at least one negative eigenvalue. On the other side, all  stable background  configurations necessarily  must have zero field strength, \textit{i.e.} $\lambda_{\Gamma}=0$.

Notice that such result depends on the choice of the gauge group ($SU(N)$) and of the number of dimensions of the torus ($T^2$). 

Change for example  the gauge group, considering  instead $U(N)$ on a $T^2$. $U(N)$ is a non-simple group and, as we have discussed before,   it is possible to consider solutions of the equations of motion with non-trivial field strength pointing to the internal direction associated to the identity. In this case the background covariant derivatives  defined in eq.~(\ref{dev_cov}), reduce to the ordinary ones and consequently  they commute. The operators $\mathcal{M}^2_{z \overline{z}}$, $\mathcal{M}^2_{\overline{z} z}$  are then given by the expressions in eq.(\ref{eig_semp}) and, therefore, are semi-positive defined. In this case, it is, therefore, possible to have stable background with  non-trivial field strength. Notice that these stable configurations have non-zero energy and are classified   by some non-trivial topological charge: in this case the first Chern class. 

Change now, instead, the number of dimensions of the torus. Consider for example $SU(N)$ on $T^4$ ($n=2$). In this case, eq.~(\ref{massa_2_0}) reduces to
\be
\mathcal{M}^2_{z_1 \overline{z}_1} \,|0\rangle &=& \left(\lambda^0_{\Gamma_2}  - \lambda^0_{\Gamma_1}  \right) \,|0\rangle  \nonumber \\
\mathcal{M}^2_{z_2 \overline{z}_2} \,|0\rangle &=& \left(\lambda^0_{\Gamma_1}  -  \lambda^0_{\Gamma_2}  \right) \,|0\rangle  \,. \nonumber 
\label{massa_1_0_T4}
\en 
Unlike for $SU(N)$ on $T^2$, it is possible to have non-negative eigenvalues if the relation
\be
\lambda^0_{\Gamma_1} =  \lambda^0_{\Gamma_2} 
\en 
is fulfilled.
Changing the number of torus dimensions,  stable background configurations with non-trivial field strength can thus exist \cite{thooft, van_baal}. Notice though, that although the background field strength is  non-trivial, the energy can be zero. The stable configurations with  non-zero energy are  classified by  some non zero  topological charge: in this particular case, the second Chern class. \\

\section{Coordinate dependent vs constant transition functions}
\label{section_U}

In the previous section, we have provided a novel demonstration of the fact that, on a two-dimensional torus, only non-simple gauge groups admit stable configurations with non-zero energy. In particular, for the case of $SU(N)$ on $T^2$  we have shown that all stable configurations are \textit{flat connections}, that is  configurations characterized by $\mathbf{F}_{ab}=0$ and thus zero energy. A flat connection  is a pure gauge configuration\footnote{Here we adopt the same approach and notation used in ref. \cite{ABGRS}.}  given by
\be
B_a = \frac{i}{g} U(y) \partial_a U^\dagger(y) \,.
\label{min}
\en
The problem of finding the non-trivial vacuum of the theory reduces then to build a $SU(N)$ gauge transformation $U(y)$ compatible with the periodicity conditions.
Substituting  eq.~(\ref{min}) into eq.~(\ref{boun_cond_A}), it follows that 
$U(y)$ has to satisfy  
\be
U(y+l_a) = \,\Omega_a(y) \,U(y) \,V_a^\dagger \,,
\label{U_cons}
\en 
where $\Omega_a(y)$ are the transition functions solution of eq.~(\ref{cons_cond}), while 
the $V_a$'s are constant elements of $SU(N)$ constrained by  the consistency conditions
\be
V_1 \,V_2\,=\,e^{2 \pi i \frac{m}{N}}\, V_2 \,V_1 \,.
\label{V_cons}
\en
Notice that given the transition functions $\Omega_1(y), \Omega_2(y)$, for each \textit{non-physically-equivalent} pair of $V_1, V_2$ there exists a different gauge transformation $U(y)$ satisfying eq.~(\ref{U_cons}) and, therefore, a different zero-energy background $B_a$.
Here, \textit{physically-equivalent} means that  $V_1, V_2$ and $V_1', V_2'$ are connected by a $SU(N)$ gauge transformation which leaves invariant $\Omega_1(y), \Omega_2(y)$.

In this section, we investigate  the conditions (choice of the gauge group, number of space-like dimensions) which guarantee that eq.~(\ref{U_cons}) admit \underline{\textit{always}} a solution regardless of the choice of $\Omega_a$ and $V_a$. 
We leave for the next section the task of classifying  and describing all  \textit{non-physically-equivalent} pairs of $V_a$.
 
As in the previous section, the proof will be carried through for the general case of a
Lie gauge group $\mathcal{G}$ and a $T^{2n}$ manifold.
In this case, the 't Hooft consistency conditions read
\be
\Omega_a (y + l_b) \,\Omega_b(y) \,&=& \,\mathcal{Z}_{ab} \,\Omega_b (y+l_a) \,\Omega_a (y) \,,
\label{cons_cond_bis}
\en
where $\mathcal{Z}_{ab}$ is the embedding of the identity in the gauge space, that is:
\be
\forall g \in \mathcal{G} \hspace{1cm}  \mathcal{Z}_{ab} \,g \,= \,g \,\mathcal{Z}_{ab} \,= \,g\,.
\label{hooft_gener}
\en 
Since $\Omega_a$ have to commute up to a factor that plays the role of identity, it follows from eq.~(\ref{cons_cond_bis}) that the transition functions $\Omega_a$ have to satisfy the following periodicity conditions:
\be
\Omega_a (y + l_b) \,&=& \,g_a^{(b)}(y) \,\Omega_a (y) \,, 
\label{perio_furbo}
\en
where the phases $g_a^{(b)}(y)$ are  constrained to verify
\be
g_b^{(a)\,-1}(y)\,g_a^{(b)}(y) \,&=&\, \mathcal{Z}_{ab} \,. 
\label{g_ab}
\en
\\
\centerline{
\bigbox{
\vtop{\hsize=13cm
\textit{For a gauge group $\mathcal{G}$ on a $2n$-dimensional torus, all sets of transition functions,
solutions of 't Hooft consistency conditions, are  gauge-equivalent if and only if the group $\mathcal{G}$ is $(2n-1)$-connected, i.e. the first $(2n-1)$ homotopy groups of $\mathcal{G}$ are trivial: $\Pi_i \left(\mathcal{G}\right)=0$.}}}}

\hspace*{1em}

\textit{Proof:} Let $\Omega_a, \Omega_a^0 \in \mathcal{G}$, $a=1,..., 2n$, be generic (constant or not) sets of solutions of the consistency conditions in eq.~(\ref{cons_cond_bis}). Considering $\mathcal{G}$ as a topological space, the transition functions $\Omega_a$ ($\Omega^0_a$) can be seen as $2n$ points describing a $(2n-1)$-dimensional loop $\mathcal{L}_{2n}$ ( $\mathcal{L}^0_{2n}$) in that space 
as a consequence of the constraint coming from eq.(\ref{cons_cond_bis}). 

\begin{figure}[htp]
\centering
\includegraphics[width=6cm]{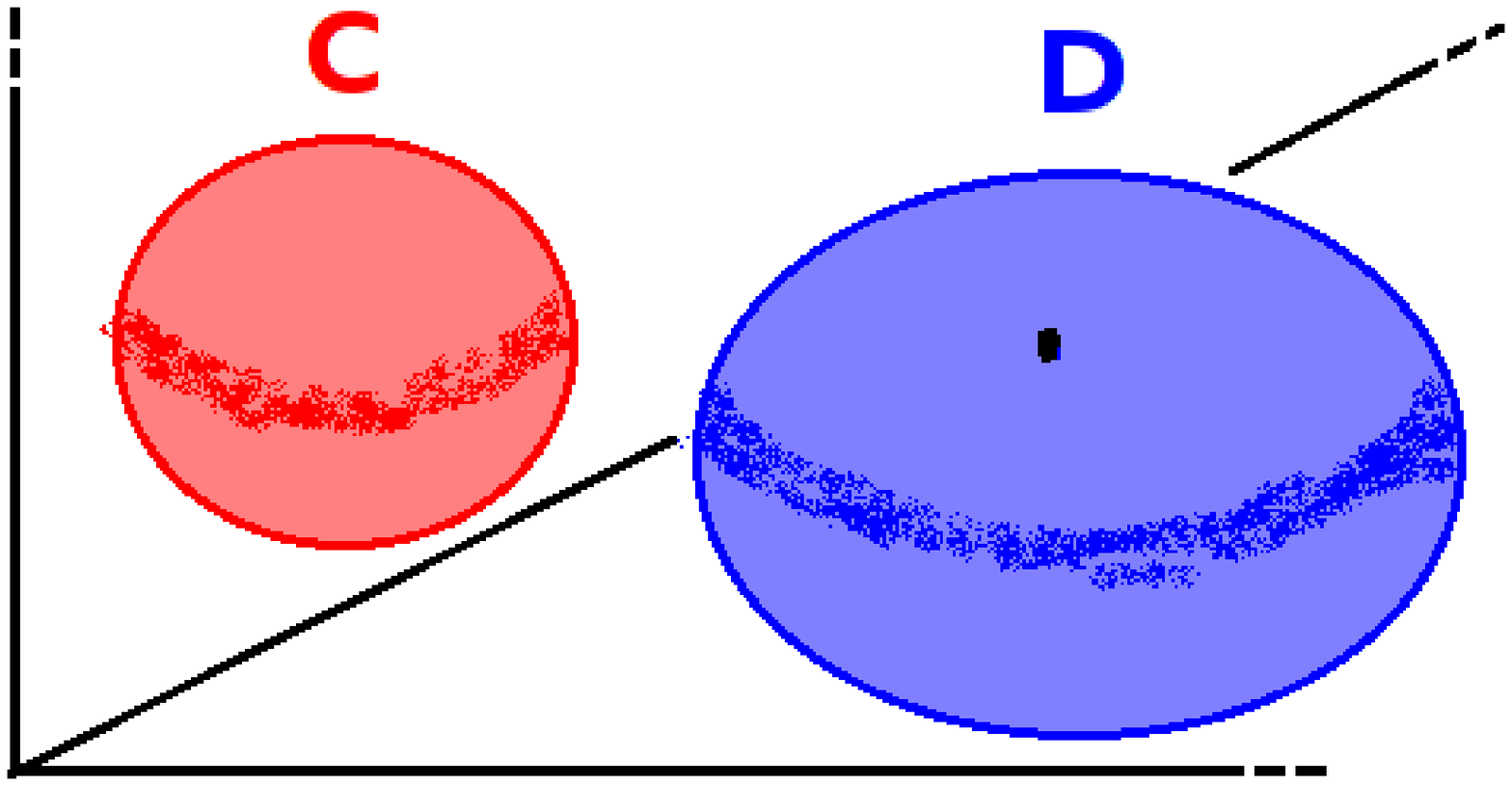}
\includegraphics[width=6cm]{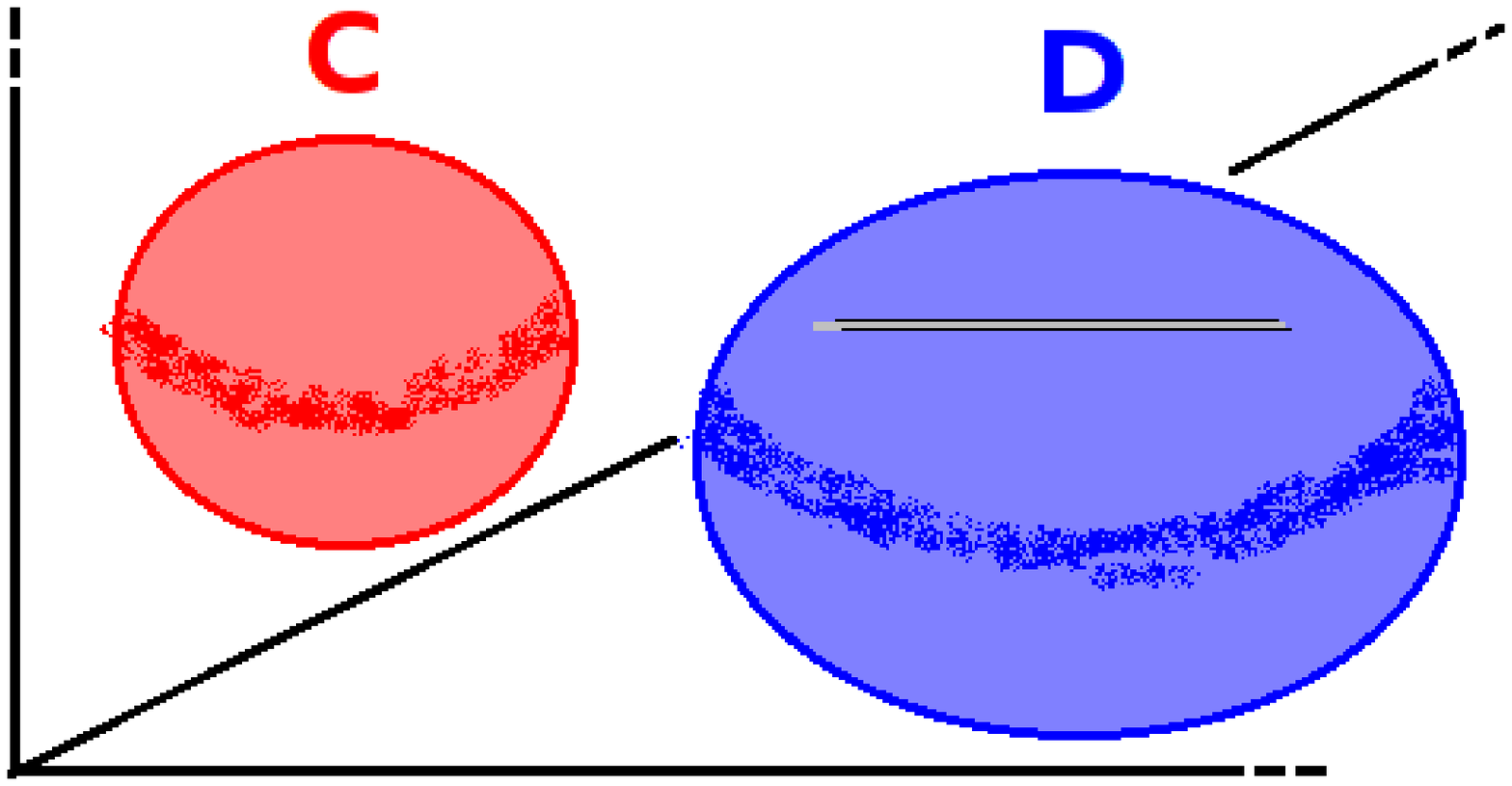}\\
\subfigure{(a) \hspace*{5cm} (b)}\\
\includegraphics[width=6cm]{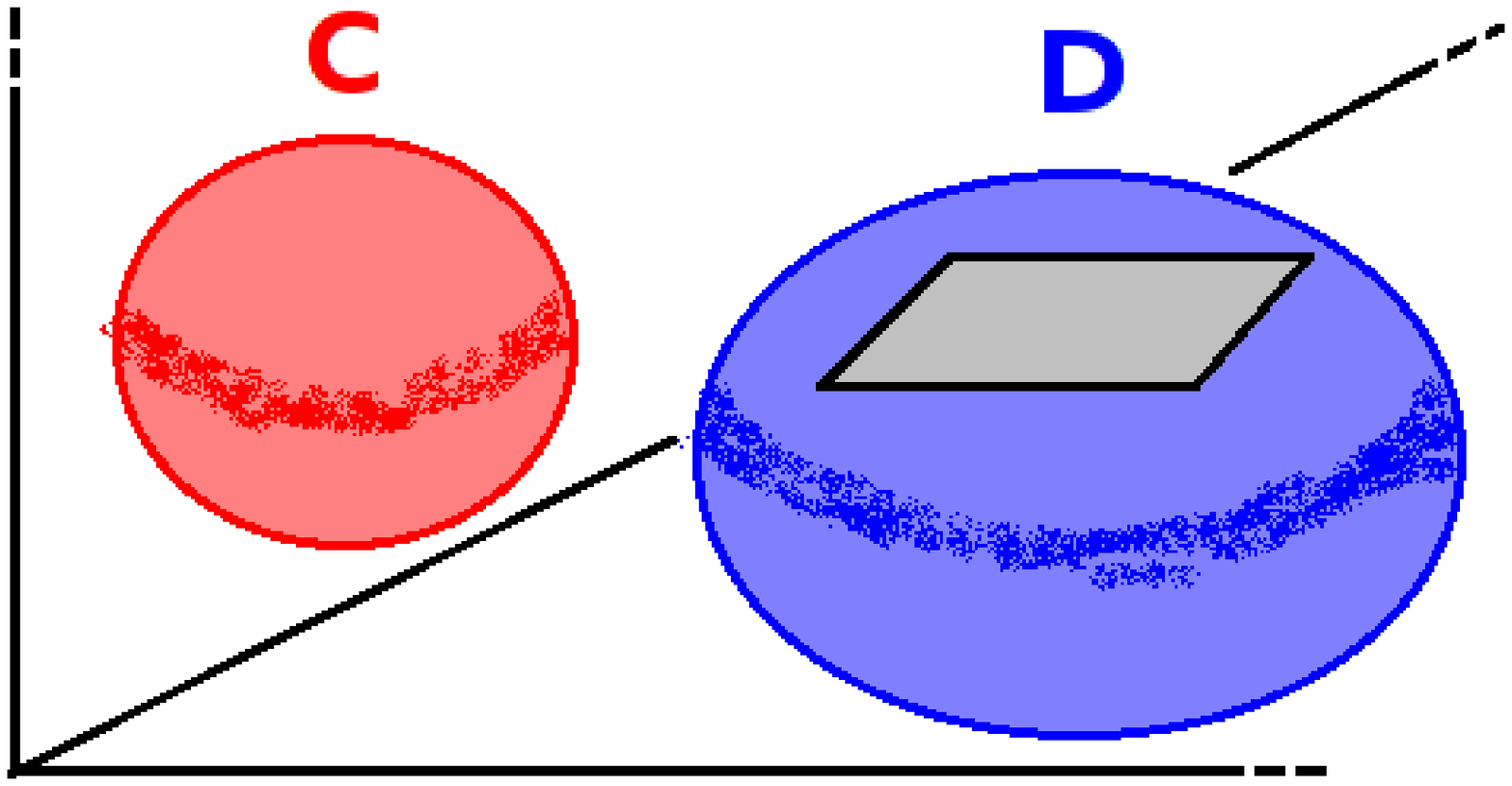}
\includegraphics[width=6cm]{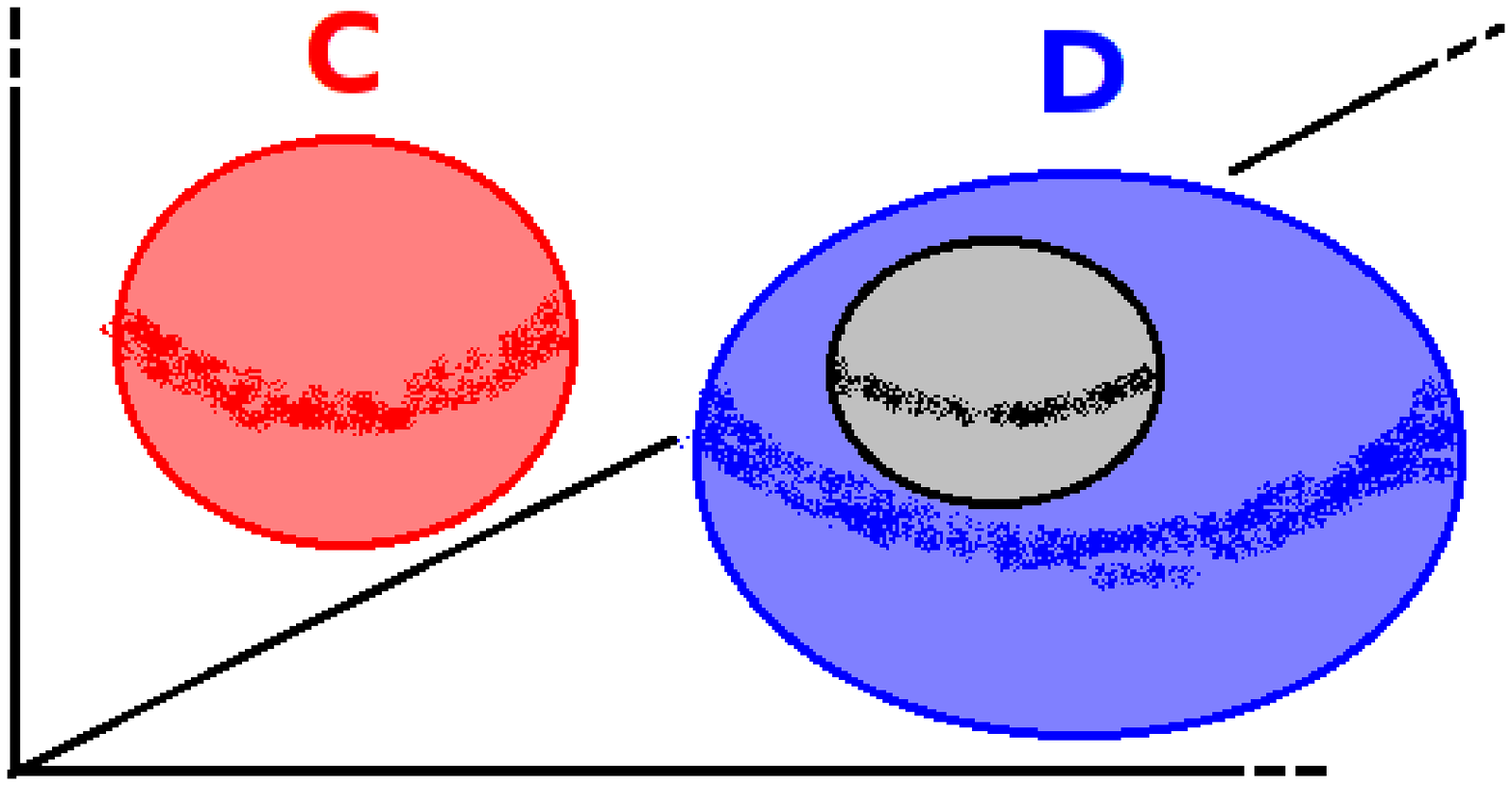}\\
\subfigure{(c) \hspace*{5cm} (d)}\\
\caption{Examples of $3$-dimensional topological spaces containing $0$, $1$, $2$ and $3-$dimensional defects (holes).  In all cases, the presence of holes avoids to obtain the $3$-dimensional loop $\mathbf{C}$ from the $3$-dimensional loop  $\mathbf{D}$ by continuous deformations. 
Notice that, although in the case (a)  \textbf{D} can be continuously deformed to a point, the latter does not belong to the space.}
\label{fig:defectos}
\end{figure}

To understand if two sets of solutions of the 't Hooft consistency conditions are gauge-equivalent, it is tantamount to determine when $\mathcal{L}_{2n}$ can be obtained from $\mathcal{L}^0_{2n}$ by a continuous deformation, \textit{i.e.} when $\mathcal{L}_{2n}$ and $\mathcal{L}^0_{2n}$ are homotopic. In particular, all  $2n$-dimensional loops, contained in a topological space, belong to the same homotopy class \textit{if and only if} they can \underline{always} be shrunk to a point, see fig. \ref{fig:defectos}. This result implies that the gauge group $\mathcal{G}$ as a topological space 
has to be $(2n-1)$-connected: $\Pi_{i}\left(\mathcal{G}\right)= 0$, $\forall i=1,...,2n-1$.

The previous reasoning can be re-formulated in a more precise way as follows. 
We consider the product of transformations changing step by step  a given $\Omega_a$ into a $\Omega^0_a$:
\be
U(y) = \,\prod_{r=1}^{2n} \,U_r  (y) \,.
\label{cazzona}
\en
By construction, therefore, $U_1(y) \in \mathcal{G}$ transforms $\Omega_1 \rightarrow \Omega_1^0$, $U_2(y) \in \mathcal{G}$ transforms $U_1(y+l_2)\,\Omega_2 U_1^\dagger(y) \rightarrow \Omega_2^0$ and leaves invariant $\Omega_1^0$, $U_3(y) \in \mathcal{G}$ transforms $U_2(y+l_3)\,U_1(y+l_3)\,\Omega_3 U^\dagger_1(y)\,U_2^\dagger(y) \rightarrow \Omega_3^0$ and leaves invariant $\Omega_1^0, \,\Omega_2^0$, etc...

Suppose that all $U_r$, with $r<\overline{r}$ and fixed $\overline{r} \in \left[ 1, 2n\right]$,  exist regardless of the choice of $\Omega_a$ and $\Omega_a^0$.
We want to show that the existence of $U_{\overline{r}}(y)$ is necessary and sufficient condition to shrink to a point all $(\overline{r}-1)$-loops of $\mathcal{G}$. 

The transformation $U_{\overline{r}}$ is defined as the transformation that allows
\be
 & U_{\overline{r}}(y) & \nonumber \\
\Omega_{\overline{r}} &\Longrightarrow & \Omega^0_{\overline{r}} \,\,,
\en
and that leaves invariant all $\Omega_r^0$ with $r<\overline{r}$.
Such a gauge transformation has to satisfy the following periodicity conditions
\be
\label{U_r_periodicity_i}
U_{\overline{r}} (y + l_r) &=& \,\Omega_r^0(y) \,U_{\overline{r}} (y)\,\Omega_r^{0\,-1}(y) \,\,,\hspace{2cm}\forall\,\,\,\,r<\overline{r} \, \\
U_{\overline{r}} (y + l_{\overline{r}}) &=& \,\Omega_{\overline{r}}^0(y) \,U_{\overline{r}} (y)\,\Omega_{\overline{r}}^{-1}(y) \,.  
 \label{U_r_periodicity_r}
\en
To simplify the notation in what follows, let us define  
\be
s &\equiv& \left\{s_1, s_2,...,s_{\overline{r}-1}\right\} \equiv\{y_{1},..,y_{\overline{r}-1}\}\,\,,\nonumber  \\
t &\equiv&\,y_{\overline{r}} \,\,,\\
u &\equiv&\{y_{\overline{r}+1},..,y_{2n}\} \,, \nonumber 
\en 
in such a way that $
y = \left\{
y_1 ,
y_2 ,
y_3 ,
... ,
y_{2n}
\right\} \,\equiv\,
\{s, t , u\}$.
In addition, we denote with $I^{\overline{r}-1}$, the $(\overline{r}-1)$-cube defined as
\be
I^{\overline{r}-1} \equiv \left\{ \left(s_1,..., s_{\overline{r}-1}\right)\,|\, 0 \leq s_i \leq l_i \,\,\left(1 \leq i \leq \overline{r}-1\right) \right\} \,,
\en
and by $\partial I^{\overline{r}-1}$  the boundary of $I^{\overline{r}-1}$, defined as
\be
\partial I^{\overline{r}-1} \,\equiv\, \left\{\left(s_1,...,s_{\overline{r}-1}\right) \in I^{\overline{r}-1} \,|\,\,\mathrm{some} \,\,\,s_i \,=\, 0 \,\,\mathrm{or}\,\, l_i  \right\}\,.
\en
A possible choice  compatible with the periodicity condition in eq.~(\ref{U_r_periodicity_r}) is 
\be
U_{\overline{r}}(s,0, u) =& 1 &\equiv \mathbf{\mathcal{C}} (s, u) \nonumber \\
\label{alpha_beta_r}
U_{\overline{r}}(s, l_{\overline{r}}, u) =&  \Omega_{\overline{r}}^0 (s, 0, u) \, \Omega_{\overline{r}}^{-1} (s, 0,u) &\equiv \mathbf{\mathcal{D}} (s,u)\,.   
\en
Using the consistency conditions in eq.~(\ref{cons_cond_bis}), the periodicity conditions in eq.~(\ref{perio_furbo}) and the constraints in eq.~(\ref{g_ab}), it is possible to prove that the choice in eq.~(\ref{alpha_beta_r}) satisfies the periodicity conditions in  eq.~(\ref{U_r_periodicity_i}). Furthermore, it is easy  to check that  for $r < \overline{r}$, it results
\be
\mathbf{\mathcal{D}} (s+l_r,\,u) \,&=& \,\Omega_r^0 \, \mathbf{\mathcal{D}} (s,\,u) \, \Omega_r^{0\,\,-1}
= \,\mathbf{\mathcal{D}} (s,\,u) \,.
\label{iau_r}
\en
$\mathbf{\mathcal{C}}(s,u)$ and $\mathbf{\mathcal{D}}(s,u)$ are two $(\overline{r}-1)$-loops $\mathcal{C},\mathcal{D}: I^{\overline{r}-1} \times T^{2n-\overline{r}} \rightarrow \mathcal{G}$ with base point $g_{\mathbf{\mathcal{C}}},\,g_{\mathbf{\mathcal{D}}}\, \in \mathcal{G}$ respectively. They map, indeed, all points of the boundary  $\partial I^{\overline{r}-1}$ into $g_{\mathbf{\mathcal{C}}},\,g_{\mathbf{\mathcal{D}}} \in \mathcal{G}$ respectively:
\be
\mathbf{\mathcal{C}} ( s|_{\partial I^{\overline{r}-1}}, u ) &=& g_{\mathbf{\mathcal{C}}} \equiv 1 \,\,,\nonumber \\
\mathbf{\mathcal{D}} ( s|_{\partial I^{\overline{r}-1}}, u ) &=& g_{\mathbf{\mathcal{D}}} \equiv    \Omega_{\overline{r}}^0 (0, 0, u) \, \Omega_{\overline{r}}^{-1} (0, 0,u) \,.
\label{mapp} 
\en
To determine the existence of  a gauge transformation $U_{\overline{r}}(y) \in \mathcal{G}$ satisfying  eq.~(\ref{alpha_beta_r}), is therefore tantamount to verify that the $(\overline{r}-1)-$loops $\mathbf{\mathcal{C}}(s,u)$ and $\mathbf{\mathcal{D}} (s, u)$ are homotopic. 
Two  $(\overline{r}-1)-$loops $\mathbf{\mathcal{C}}(s,u)$ and $\mathbf{\mathcal{D}} (s, u)$ are homotopic  (regardless of the choice of $\Omega^0_a$ and $\Omega_a$) if and only if the $(\overline{r}-1)$-th homotopic group of $\mathcal{G}$ is trivial. The existence of the transformation $U_{\overline{r}}(y)$  guarantees, therefore, that all $(\overline{r}-1)$-loops of $\mathcal{G}$ can be shrunk to a point. 

Finally, the existence of $U(y)$ defined in eq.~(\ref{cazzona}) $\forall \,\Omega_a^0$ and $\Omega_a$ is, therefore, necessary and sufficient for  $\mathcal{G}$ to be $(2n-1)$-connected.\\

Summarizing, we have shown that depending on the gauge group $\mathcal{G}$ and on the number of dimensions of the torus,  eq.~(\ref{U_cons}) may admit solution independently on the choice of $\Omega_a$ and $V_a$, satisfying eq.~(\ref{cons_cond}) and eq.~(\ref{V_cons}), respectively.
For example, for a $SU(N)$ gauge theory on a two-dimensional torus,  since  such group is simply connected (that is $\Pi_1(SU(N))=0$), two sets of solutions of the 't Hooft consistency condition are always gauge equivalent: eq.~(\ref{U_cons}) always admits a solution.

If we increase the number of dimensions of the torus or change the gauge group, this result does not remain  necessarily valid. For example: 
\begin{itemize}
\item $SU(N)$ is not $3-$connected, since $\Pi_3\left(SU(N)\right) = \mathbf{Z}$. In consequence, if we consider $SU(N)$ on $T^4$ not all the sets of transition functions can be related by a gauge transformation.
\item $U(N)$  is not simply connected and then, for $U(N)$ on $T^2$, there exist solutions of the consistency conditions in eq.~(\ref{cons_cond_bis}), inequivalent to the constant ones.
\end{itemize}

\section{Vacuum symmetries and 4D spectrum}
\label{section_back}

In this section, we want to find and to catalogue the possible different classical vacua for a $SU(N)$ theory on a $T^2$, to discuss their  symmetries  and to compute the effective four-dimensional  spectrum of  fluctuations $\{A_\mu, A_a\}$.

In a general background gauge, such exercise can turn out to be very complicate since  we have at the same time non-trivial transition functions $\Omega_a$ and non-trivial vacuum gauge configuration $B_a$. To simplify the discussion, it is useful to introduce the background \textit{symmetric gauge}: the gauge in which $B^{sym}_a =0$ and $\Omega_a^{sym} = V_a$. 

To determine whether it is possible to go in the background \textit{symmetric gauge} translates in to solve eq.(\ref{U_cons}) and therefore the $SU(N)$ gauge transformation $S(y)$ that allows to go in that gauge is simply
$S(y)= U^\dagger(y)$.

In the background \textit{symmetric gauge}, to classify the classical vacua means to find all possible constant transitions functions $\Omega_a^{sym}=V_a$, solutions of 't Hooft consistency conditions. To determine the residual symmetries  reduces to establish the symmetries of  $\Omega_a^{sym}= V_a$. In this gauge, in fact, the periodicity conditions for the fluctuations fields $A_M = \{A_\mu, A_a\}$ are given by
\be
A_M (y+l_a) = V_a \,A_M(y) V_a^\dagger \,,
\label{perio_costanti}
\en
and, therefore, the residual symmetries  are  associated to the $SU(N)$ generators that commute with $V_a$.

We divide our analysis in two cases: 
\begin{itemize}
\item Trivial 't Hooft non-abelian flux: $m=0$.
\item Non-trivial 't Hooft non-abelian flux: $m\neq0$.
\end{itemize}

\subsection{Trivial 't Hooft flux: $m=0$}

For $m=0$, the transition functions commute and all the classical vacuum configurations are degenerate in energy with the trivial $SU(N)$ symmetric vacuum. 
$V_a$ can be parametrized as 
\be
V_a= e^{2 \pi i \alpha_a^j H_j} \,,
\label{V_a_m=0}
\en
where $H_j$ are the $N-1$ generators of the Cartan subalgebra of $SU(N)$.  
$V_a$, and therefore the vacua, are characterized by $2 (N-1)$ real continuous parameters $\alpha_a^j$, $0\leq \alpha_a^j < 1$. $\alpha_a^i$ are non-integrable phases, which arise only in a topologically non-trivial space and cannot  be gauged-away.
Their values must be dynamically determined at the quantum level: only at this level the degeneracy among the infinity of classical vacua is removed \cite{Hosotani}. 

The solution with $\alpha_a^j=0$ is the fully symmetric one.  For $\alpha_a^i\neq 0$, the residual gauge symmetries are those associated with the generators that 
commute with $V_a$. As $V_1$ and $V_2$ commute, the symmetry breaking is rank-preserving and the maximal symmetry breaking pattern that can be achieved is 
$SU(N) \longrightarrow U(1)^{N-1}$.

The spectrum of the fluctuations reflects the symmetry breaking pattern and it is a function of the non-integrable phases. 
To give an explicit expression of that spectrum, it is useful to use the Cartan-Weyl basis for the $SU(N)$ generators.
In addition to the generators of the Cartan subalgebra $H_j$ with $j=1,..,N-1$ that satisfy
\be
\left[ H_{j_1}, H_{j_2} \right] = 0 \,\,\,\,\,\,\,\forall j_1,j_2 = 1,...,N-1 \,,
\en
we denote as $E_r$, $r=1,...,N^2-N$, all other $SU(N)$ generators such that
\be
\left[ H_{j}, E_r \right] = q^j_r E_r \,\,\,\,\,\,\,\forall \,j=1,...,N-1 \,\,\,\,\mathrm{and}\,\,\,\,\forall \,r=1,...,N^2-N\,.
\en 
In this basis, the four-dimensional mass spectrum for a gauge field $A_M^j$ belonging to the Cartan subalgebra, is the ordinary Kaluza-Klein (KK) spectrum
\be
m^2_{(j)} = 4 \pi^2 \left[ \frac{n_1^2}{l_1^2} +  \frac{n_2^2}{l_2^2} \right] \,,\hspace*{2cm} n_1, n_2 \,\,\in \,\mathbf{Z}\,.
\label{KK}
\en
For a gauge field $A_M^r$ associated to the generator $E_r$, the mass spectrum reads 
\be
m^2_{(r)} = 4 \pi^2 \left[ \left(n_1 + \sum_{j=1}^{N-1} \,q^j_r \alpha_1^j\,\right)^2 \frac{1}{l_1^2}+ \left(n_2 + \sum_{j=1}^{N-1} \,q^j_r \alpha_2^j\,\right)^2 \frac{1}{l_2^2}\right] \,.
\label{SS}
\en
For all $\alpha_a^j\neq 0$, the only  four-dimensional gauge fields  that continue to be massless are the $N-1$ fields belonging to the Cartan subalgebra. The spectrum shows the expected maximal symmetry breaking pattern: $SU(N) \rightarrow U(1)^{N-1}$. Finally, notice that the classical spectra described by eqs.(\ref{KK})-(\ref{SS}) depend on the gauge indices but do not depend on the Lorentz ones: from the four-dimensional point of view, the scalars and the gauge bosons coming from internal and ordinary components of a higher-dimensional gauge field, respectively, are expected to be degenerate at least at the classical level. However, such degeneracy is always removed at the quantum level \cite{Hernandez}. 

\subsection{Non-trivial 't Hooft flux: $m\neq0$}

For $m \neq0$, the transition functions do not commute and all stable vacuum configurations induce some symmetry breaking. 
Eq.~(\ref{V_cons}) reduces to the so-called two-dimensional twist algebra \cite{'tHooft:1979uj}. The first solutions  were found in Refs.\cite{Groeneveld:1980tt,Ambjorn:1980sm}. The problem for the $4$-dimensional case was addressed and solved in Refs.\cite{thooft,van_baal,Gonzalez-Arroyo:1982hz,Gonzalez-Arroyo:1982ub}. The most general  solution up to 4 dimensions was obtained in Refs.\cite{vanBaal:1983eq,Brihaye:1983yd} and the $d$-dimensional case was studied in Refs.\cite{vanBaal:1985na,vanGeemen:1984wy,tony}. 

In order to analyze the possible constant transition functions which are solutions of the two-dimensional twist algebra, let us define the quantity\footnote{g.c.d.= great common divisor.} 
\be
\mathcal{K} \equiv g.c.d.\left( m, N\right),
\en
and divide the analysis in two cases: $\mathcal{K} = 1$ and $\mathcal{K} >1$.

\subsubsection*{\textbf{$\mathcal{K}=1$}}

In this case, 
the possible solutions of eq.(\ref{V_cons}) are of the type
\cite{thooft} 
\be
\left\{\begin{array}{ccc}
V_1 & = & P_{(N)}^{\alpha_1}\,Q_{(N)}^{\beta_1} \\
V_2 & = & P_{(N)}^{\alpha_2} \,Q_{(N)}^{\beta_2}  
\end{array} \right. \,,
\label{choice}
\en
where the constant $N \times N$ matrices $P_{(N)}$ and $Q_{(N)}$ are defined as 
\be
\left\{ \begin{array}{lcr}
\left(P_{(N)}\right)_{kj} & = & e^{- 2 \pi i \frac{(k-1)}{N}} \,\,e^{i \pi \frac{N-1}{N}}\,\, \delta_{kj}\\
\left(Q_{(N)}\right)_{kj} & = & e^{i \pi \frac{N-1}{N}} \,\, \delta_{k,j-1}
\end{array} \right. \,,
\label{PQ_def}
\en
and satisfy the conditions 
\be
\left(P_{(N)}\right)^N &=& \left(Q_{(N)}\right)^N  =  e^{\pi i (N-1)} \nonumber \\ 
P_{(N)}\,Q_{(N)} &=& e^{\frac{2 \pi i}{N}} Q_{(N)}P_{(N)} \,.
\label{PQ_properties}
\en
For $m\neq0$ and $\mathcal{K} =1$ we have a finite number of matrices of the type in eq.(\ref{choice}) characterized by discrete parameters $\alpha_1,\,\alpha_2,\,\beta_1,\,\beta_2\in \left[0, N-1\right]$ (modulo $N$), which have to satisfy the consistency condition 
\be
\alpha_1 \,\beta_2 \,\,-\,\,\alpha_2\,\beta_1\,\,=\,\,\mathrm{Det}\left(\begin{array}{cc}
\alpha_1 & \alpha_2 \\
\beta_1 & \beta_2
\end{array}\right)\,\,=\,\,\mathrm{Det}M\,\,=\,\,m \,.
\label{cons_alp_beta_m_aa}
\en
Notice that  $\alpha_1$, $\alpha_2$, $\beta_1$ and $\beta_2$ cannot be simultaneously zero.
The condition in eq.(\ref{cons_alp_beta_m_aa}) is invariant under the following biunimodular transformation
\be
M \rightarrow M' \equiv U M V \,,
\label{biuni_cap}
\en
where $U,V \in SL(2,\mathbb{Z})$. The invariance of the condition in eq.(\ref{cons_alp_beta_m_aa}) under the biunimodular transformations of eq.(\ref{biuni_cap}) 
implies that all different matrices in eq.(\ref{choice}) are different parametrizations of the same vacuum. The only physical parameter is the value of $\mathcal{K}$.
 
The constant matrices in eq.(\ref{choice}) form an irreducible representation of the two-dimensional twist algebra: they have $N$ different eigenvalues. An irreducible representation is unique modulo gauge transformations in eq.(\ref{omega_gauge}), and multiplication of the matrices by a constant in eq.(\ref{omega_centro}).

Now, we want to show that the periodicity conditions in eq.(\ref{perio_costanti}) with the choice in eq.(\ref{choice}) completely break  the original $SU(N)$ symmetry group.
To prove this statement, we introduce the following basis for the generators of 
$SU(N)$:
\be
\tau^{(N)}(\Delta,k_{\Delta}) &=& \sum_{n=1}^{N} \, e^{2 \pi i \frac{n}{N}\,k_\Delta}\,\lambda^{(N)}_{(n,n+\Delta)}  \,,
\label{gen_nodiag}
\en 
where $\Delta=0,1,...,N-1$, and $k_{\Delta=0}\equiv k_0 =1,...,N-1$ and $k_{\Delta\neq0}=0,1,...,N-1$. The matrices $\lambda^{(N)}_{(n,m)}$ are the $N\times N$ matrices defined by 
\be
\left(\lambda^{(N)}_{(n,m)}\right)_{ij} \equiv  \delta_{ni} \delta_{mj} \,.
\label{lambda_ij}
\en 
The matrices $\tau^{(N)}(\Delta,k_\Delta)$ are eigenstates of the operators $P_{(N)},\,Q_{(N)}$  with eigenvalues $ e^{2 \pi i \frac{\Delta}{N}}$ and $e^{2 \pi i \frac{k_\Delta}{N}}$, respectively and satisfy the following properties
\be
\nonumber &&\hspace*{-0.8cm} 
\mathrm{Tr} \,\,\tau^{(N)}(\Delta,k_\Delta) \,\,\,= 0  \\
\nonumber && \hspace*{-0.8cm} 
\frac{1}{N} \,\,\mathrm{Tr} \left[\,\tau^{(N)\,\dagger}(\Delta,k_\Delta) \tau^{(N)}(\Delta',k_\Delta')\right] \,\,= \,\,\delta_{\Delta,\Delta'}\,\delta_{k_\Delta,k_\Delta'}
\\
\nonumber &&\hspace*{-0.8cm} 
\left[ \tau^{(N)} \left(\Delta,\,k_\Delta \right),\,\tau^{(N)}\left(\Delta',\,k_\Delta'\right)\right] \,= \,\left(e^{\frac{2 \pi i}{N}\,\Delta\,k_\Delta'}\,-\,e^{\frac{2 \pi i}{N}\,\Delta'\,k_\Delta}\right)\,\tau^{(N)}\left(\Delta\,+\,\Delta',\,k_\Delta + k_\Delta'\right) 
 \,.
\en
In this basis, the periodicity conditions reads
\be
V_a\, \tau^{(N)}(\Delta,k_\Delta) \,V_a^\dagger \,\,&=&\,\, e^{2 \pi i \frac{\alpha_a\,\Delta\,\,+\beta_a\,k_\Delta}{N}}\,\, \tau^{(N)} (\Delta,k_\Delta)  \hspace{0.8cm} \mathrm{for} \,\,\,\,\,a=1,2 \,\,.
\label{cho}
\en
The residual symmetries are associated to $SU(N)$ generators that commute simultaneously with $V_1$ and $V_2$, that is to those $\tau^{(N)} (\Delta,k_\Delta)$ for which it results
\be
\frac{\alpha_a\,\Delta\,\,+\beta_a\,k_\Delta}{N} \,&\in&\, \mathbf{Z} \,\,.
\label{straight}
\en
Using the Bezout theorem, it is possible to prove that the number of $SU(N)$ generators $\tau^{(N)} \left(\Delta,k_\Delta\right)$ satisfying the condition in eq.~(\ref{straight}), that is the dimension of the residual symmetry group $\mathcal{H} \subset SU(N)$, results
\be
\mathrm{Dim}\left[\mathcal{H}\right]\,\,=\,\,\mathcal{K}^2 -1 \,.
\label{dim_ropt_m}
\en 
For $\mathcal{K}=1$, therefore, $SU(N)$ is completely broken by the $V_a$ in eq.(\ref{choice}). 

\subsubsection*{\textbf{$\mathcal{K}>1$}}

In this case, the matrices in eq.(\ref{choice}) with the constraint in eq.(\ref{cons_alp_beta_m_aa}) form a reducible representation of the twist algebra.
Unlike the $\mathcal{K} =1$ case, indeed, the matrices in eq.(\ref{choice})
are now invariant under a subgroup $\mathcal{H}$ of $SU(N)$ given by
\be
\mathcal{H}\,=\,SU(\mathcal{K}) \,\subset\,SU(N)\,.
\label{ropt_m}
\en
Since $\mathcal{K} < N$, the rank of $\mathcal{H}$ is always less than the one of $SU(N)$.

For $\mathcal{K}>1$, the irreducible representations of the twist algebra (up to gauge transformations in eq.(\ref{omega_gauge}) and multiplications by a constant as in eq.(\ref{omega_centro})) can be obtained in the following way
\be
V_1 &=& \omega_1 \,\,P^{\alpha_1}\,Q^{\beta_1} \nonumber \\
V_2 &=& \omega_2 \,\,P^{\alpha_2}\,Q^{\beta_2} \,.
\label{gaiarda}
\en
Now, the $SU(N)$ constant matrices $P$ and $Q$ are given by
\be
\nonumber
P & \equiv & \left( 
\begin{array}{cccc}
P_{(N/\mathcal{K})} & 0 & ...& 0 \\
0 & P_{(N/\mathcal{K})} & ...& 0 \\
... & ... & ... & ...  \\
0 & 0 & ... & P_{(N/\mathcal{K})}
\end{array}
\right)_{(\mathcal{K}\times \mathcal{K})} \\
Q & \equiv & \left( 
\begin{array}{cccc}
Q_{(N/\mathcal{K})} & 0 & ...& 0 \\
0 & Q_{(N/\mathcal{K})} & ...& 0 \\
... & ... & ... & ...  \\
0 & 0 & ... & Q_{(N/\mathcal{K})}
\end{array}
\right)_{(\mathcal{K}\times \mathcal{K})} \,\,,
\label{PQ_paraculi}
\en
where $P_{N/\mathcal{K}}$ and $Q_{N/\mathcal{K}}$ are $N/\mathcal{K} \times N/\mathcal{K}$  matrices  defined as in eqs.(\ref{PQ_def})-(\ref{PQ_properties}) with the change $N \rightarrow N/\mathcal{K}$. In this case, the discrete parameters $\alpha_1$, $\alpha_2$, $\beta_1$, $\beta_2$ have to satisfy the constraint in eq.(\ref{cons_alp_beta_m_aa}) where $m \rightarrow m/\mathcal{K}$. The constant matrices $\omega_1,\omega_2$ are elements of the subgroup  $SU(\mathcal{K}) \subset SU(N)$ and have to satisfy the constraint
\be
\omega_1 \,\omega_2\,\,= \,\,\omega_2\,\omega_1 \,.
\en
As in the $m=0$ case, $\omega_1$ and $\omega_2$ commute and therefore they can be parametrized in terms of generators ($H_j$) belonging to the Cartan subalgebra of $SU(\mathcal{K})$:
\be
\omega_a \,=\,e^{2 \pi i\, \,\sum_{j=1}^{\mathcal{K}-1}\,\phi_a^j \,H_j} \,.
\label{omega_piccola}
\en
$\phi_a^j$ are $2 (\mathcal{K}-1)$ real continuous parameters taking values in the interval $0 \leq \phi_a^j < 1$ (modulo integers). They are non-integrable phases and their values must be dynamically determined at the quantum level: only at this level the degeneracy among the infinity of classical values can be removed \cite{Hernandez}.

\vspace*{1cm}

Summarizing, in the $m\neq0$ case  the possible solutions of condition in eq.(\ref{V_cons}) and therefore the possible stable vacua are given by the formula in eq.(\ref{gaiarda}) with the restriction that for $\mathcal{K}=1$ it results $\omega_1=\omega_2=1$. Such vacua are characterized by the value of $\mathcal{K}$ and by $2(\mathcal{K} - 1)$ continous  $\phi_a^j$ parameters. 
\begin{enumerate}
\item In the case $\phi_a^j = 0$ $\forall j$, the residual symmetry group is given by the $SU(N)$ subgroup that commutes with the matrices $P$ and $Q$ in eq.(\ref{gaiarda}). These matrices always induce some degree of symmetry breaking due to their non-trivial commutation rules.  Such symmetry breaking mechanism is rank-lowering and, as discussed before, the pattern that can be achieved results $SU(N) \rightarrow SU(\mathcal{K})$. 
In particular, for $\mathcal{K}=1$, $SU(N)$ is completely broken.
\item For $\mathcal{K}>1$ and non-vanishing phases $\phi_a^j$, there is additional degrees of freedom  with respect to  the symmetry breaking pattern. Since  $\omega_1$ and $\omega_2$ commute, such new symmetry breaking mechanism preserves the rank of  $SU(\mathcal{K})$, and the maximal symmetry breaking that can be achieved is  
\be
SU(N) \rightarrow SU(\mathcal{K}) \rightarrow U(1)^{\mathcal{K} -1} \,.
\label{symm_break_mneq1}
\en
\end{enumerate}
In order to discuss the effective four dimensional mass spectrum, we need to diagonalize the constant transition functions $V_a$ and consequently the periodicity conditions. To do that, we generalize the basis of  generators of $SU(N)$ introduced in eq.(\ref{gen_nodiag}) as follows
\be
\tau_{(\rho,\sigma)} (\Delta, k_\Delta) &=& \left\{ 
\begin{array}{lll}
\mathrm{if}\, \left\{\begin{array}{l}
\rho=\sigma \\
\Delta=k_\Delta=0
\end{array} \right.
 & \Rightarrow &\left(\sum_{i=1}^{\rho} \lambda^{(\mathcal{K})}_{(i,i)} - \rho \lambda^{(\mathcal{K})}_{(\rho+1,\rho+1)}\right) \otimes 1_{(N/\mathcal{K})} \\ 
&\\
\mathrm{else} & \Rightarrow & \lambda^{(\mathcal{K})}_{(\rho,\sigma)} \otimes \tau^{(N/\mathcal{K})} (\Delta, k_\Delta) 
\end{array}
\right.
\label{base_fighetta}
\en
where now $\Delta,\,k_\Delta=0,..,N/\mathcal{K}-1$ and $\rho,\sigma=1,...,\mathcal{K}$ excluding the case $\Delta=k_\Delta=0$ and $\rho=\sigma$ in which $\rho=1,..., \mathcal{K}-1$. Here, $\tau^{(N/\mathcal{K})} (\Delta, k_\Delta)$ are $N/\mathcal{K} \times N/\mathcal{K}$ matrices defined as in eq.(\ref{gen_nodiag}), but with the change $N\rightarrow N/\mathcal{K}$. In the same way, the $\mathcal{K} \times \mathcal{K}$ matrices $\lambda^{(\mathcal{K})}_{(\rho,\sigma)}$ can be obtained from eq.(\ref{lambda_ij}) with the substitution $N \rightarrow \mathcal{K}$.

In this basis, the generators which commute with $P$ and $Q$ and form the subgroup $SU(\mathcal{K})$, can be identified with $\tau_{\rho,\sigma} (0,0)$. In particular,  the generators belonging to the Cartan subalgebra of $SU(\mathcal{K})$ and appearing in eq.(\ref{omega_piccola}) are given by $H^{j\equiv\rho}=\tau_{\rho,\rho}(0,0)$.

Now, the periodicity condition read
\be
V_a\, \tau_{(\rho,\sigma)}(\Delta,k_\Delta) \,V_a^\dagger \,\,&=&\,\, e^{2 \pi i \left(\frac{\mathcal{K}}{N}(\alpha_a\,\Delta\,\,+\beta_a\,k_\Delta)\,\,+\phi_a^{(\rho)}\,\,-\phi_a^{(\sigma)}\right)}\,\, \tau_{(\rho,\sigma)} (\Delta,k_\Delta)  \,\,.
\label{cho_K}
\en
At the classical level, the effective four-dimensional mass spectrum is, also in this case, independent of the Lorentz index $M$  and takes the following form
\be
\label{m_spectrum_dani}
m^2 = 4 \pi^2 \,\sum_{a=1}^2\,\, \left(n_a + \frac{\mathcal{K}}{N}\,(\alpha_a\,\Delta\,\,+\,\,\beta_a\,k_\Delta )\,\,+\,\,\phi_a^{(\rho)}-\phi_a^{(\sigma)} \right)^2 \frac{1}{l_a^2}  \,,
\en
with $n_1, n_2 \in \mathbf{Z}$.
 
The effective four-dimensional mass spectrum reflects the symmetry breaking pattern discussed before. Given $\alpha_1,\,\alpha_2,\,\beta_1,\,\beta_2$ and for all $\phi_a^{\rho}=0$, only 
gauge bosons associated to the generators $\tau_{(\rho,\sigma)} (0,0)$ of $SU(\mathcal{K})$ admit zero modes. 
Since $\alpha_1,\,\alpha_2,\,\beta_1,\,\beta_2$ cannot be  simultaneously zero, the spectrum described by eq.~(\ref{m_spectrum_dani}) exhibits always some degree of symmetry breaking. For $\mathcal{K}>1$ and all $\phi_a^j\neq0$, the only massless modes arise from gauge bosons associated  to the Cartan subalgebra of $SU(\mathcal{K})$: $\tau_{(\rho,\rho)} (0,0)$.\\


Finally, it is worth to  underline the different nature of  the symmetry breaking for  the two cases of trivial ($m=0$) and non-trivial ($m\neq0$) 't Hooft non-abelian flux. In the $m=0$ case, the gauge symmetry breaking mechanism is, indeed,  the Hosotani mechanism \cite{Hosotani}: it is always possible to choose an appropriate background gauge, compatible with the consistency conditions, in which the transition functions are trivial ($V_1=V_2=\mathbf{1}$) and some extra space-like components of the six-dimensional gauge fields $\mathbf{A}_a$ acquire a vacuum expectation value (VEV): $\langle\mathbf{A}_a\rangle=B_a$. In this case, the symmetry breaking can be seen as spontaneous in the following sense:
\begin{enumerate}
\item For each $4$-dimensional massive gauge field $\mathbf{A}_{\mu}$, there exists a linear combination of the $\mathbf{A}_a$ that play the role of a $4$-dimensional scalar pseudo-goldstone boson, eaten by the $4$-dimensional gauge bosons to become a longitudinal gauge degree of freedom. 
\item The VEV of $\mathbf{A}_a$ works as the order parameter of the symmetry breaking mechanism. In particular, it is possible  to deform $\langle\mathbf{A}_a\rangle $ to zero compatibly with the consistency conditions, so as to restore all the initial symmetries. 
\end{enumerate}
In the $m \neq 0$ case, we cannot interpret all steps of  symmetry breaking in eq.(\ref{symm_break_mneq1}) as spontaneous: 
\begin{enumerate}
\item The $SU(N) \rightarrow SU(\mathcal{K})$ symmetry breaking is due to the constant matrices $P$ and $Q$ of eqs.(\ref{gaiarda})-(\ref{PQ_paraculi}). They don't commute as a consequence of the non-trivial value of the 't Hooft flux. 
\textit{The symmetry breaking can not be related \textbf{only} to the VEV of} $\mathbf{A}_a$. Although for each massive $4$-dimensional gauge boson $\mathbf{A}_{\mu} \in \frac{SU(N)}{SU(\mathcal{K})}$ there exists a $4$-dimensional pseudo-goldstone boson, in this case it is not possible to determine an order parameter that can be  deformed compatibly with the consistency conditions in such a way to restore all the initial symmetries.
\item The $SU(\mathcal{K}) \rightarrow U(1)^{\mathcal{K} -1} $ symmetry breaking is due to the matrices $\omega_1$ and $\omega_2$ that commute. This symmetry breaking mechanism exactly works as in the $m=0$ case: it is always possible to choose an appropriate $SU(\mathcal{K})$ background gauge in which $\omega_1=\omega_2=\mathbf{1}$ and some extra space-like components of the six-dimensional gauge fields $\mathbf{A}_a \in SU(\mathcal{K})$ acquire a VEV.
This step can be understood  as a consequence of a spontaneous symmetry breaking mechanism.
\end{enumerate}

\section{Conclusions}
\label{conclusioni}

We have studied extra-dimensional gauge theories with the extra dimensions compactified \textit{\'a la} Scherk-Schwarz on toroidal manifolds. 

Using the analogy with the harmonic oscillator, we have analyzed the vacuum energy for a general  group on an even-dimensional torus. For the particular case of $SU(N)$ on $T^2$, we have re-obtained the  well-known result that all stable vacua compatible with four-dimensional Poincar\'e invariance and zero four-dimensional instanton number have zero energy. 

We have, then, studied the classical zero-energy vacua, for a gauge theory  on an even-dimensional torus, with  periodicity conditions  satisfying the 't  Hooft consistency conditions. In the $SU(N)$ on $T^2$ case, the set of degenerate and inequivalent classical zero-energy vacua is completely determined by the set of \textit{constant} transition functions $V_a$ solutions of the 't Hooft consistency conditions. 
We have explicitly proved that such result depends on the particular choice of the gauge group and of the number of extra dimensions.

The number of vacua, the residual symmetries and the nature of the symmetry breaking mechanism are determined by the value of the 't Hooft non-abelian flux:
\begin{itemize}
\item For trivial 't Hooft flux, $m=0$, it results a continuum of vacua, degenerate at the classical level with the $SU(N)$ symmetric one. The symmetry breaking is rank-preserving and spontaneous since it is exactly as the Hosotani mechanism.  
\item The main novel result of this paper is the explicit demonstration of the symmetry breaking pattern and  the four-dimensional mass spectrum for the case of non-trivial 't Hooft flux. For $m\neq0$, the number of vacua depends on the value of $\mathcal{K}=\mathrm{g.c.d.}(m,N)$. For $\mathcal{K}=1$ there exists only one classical vacuum and $SU(N)$ is completely broken. For $\mathcal{K}>1$, there is a degeneracy among an infinity of classical vacua and  $SU(N)$ is partially broken in all of them.
In particular, the symmetry breaking can be seen as due to two different mechanisms. The first mechanism induces  a rank-lowering symmetry breaking that, from four-dimensional point of view, can be understood as explicit. The second mechanism, instead, gives rise to a rank-preserving symmetry breaking that can be interpreted as spontaneous.
\end{itemize}

\section*{Acknowledgements}

I acknowledge for very interesting discussions J. Bellorin,  C. Biggio, A. Broncano,  E. Fernandez-Martinez, M. Garc\'ia Per\'ez, D. Hernandez. I would like to thank in a special way J.Alfaro for suggesting to me how to face the problem discussed in section 3. I am grateful to  G. von Gersdorff for a very useful e-mail correspondence. I am specially indebted for discussions and for reading the manuscript to M.B. Gavela and S. Rigolin.  I also acknowledge MECD for financial support through FPU fellowship AP2003-1540.


\addcontentsline{toc}{chapter}{Bibliography}

\begin{thebibliography}{9}

\bibitem{ss}
J.~Scherk and J.~H.~Schwarz,
Nucl.\ Phys.\ B {\bf 153}, 61 (1979);
Phys.\ Lett.\ B {\bf 82}, 60 (1979).

\bibitem{early}
N.~S.~Manton,
Nucl.\ Phys.\ B {\bf 158} (1979) 141;
D.~B.~Fairlie,
Phys.\ Lett.\ B {\bf 82} (1979) 97;
J.\ Phys.\ G {\bf 5} (1979) L55;
P.~Forgacs, N.~S.~Manton,
Commun.\ Math.\ Phys.\  {\bf 72} (1980) 15;
N.~V.~Krasnikov,
Phys.\ Lett.\ B {\bf 273} (1991) 246;

\bibitem{5Drecenti}
G.~von Gersdorff, N.~Irges, M.~Quiros,
Nucl.\ Phys.\ B {\bf 635} (2002) 127
[hep-th/0204223];
G.~Burdman, Y.~Nomura,
Nucl.\ Phys.\ B {\bf 656} (2003) 3
[hep-ph/0210257].
C.~A.~Scrucca, M.~Serone and L.~Silvestrini,
Nucl.\ Phys.\ B {\bf 669} (2003) 128
[hep-ph/0304220].
Y.~Hosotani, S.~Noda and K.~Takenaga,
Phys.\ Lett.\ B {\bf 607} (2005) 276
[hep-ph/0410193];
N.~Haba, Y.~Hosotani, Y.~Kawamura and T.~Yamashita,
Phys.\ Rev.\ D {\bf 70} (2004) 015010
[hep-ph/0401183];
N.~Haba and T.~Yamashita,
JHEP {\bf 0402} (2004) 059
[hep-ph/0401185];
N.~Haba, K.~Takenaga and T.~Yamashita,
hep-ph/0411250;
G.~Martinelli, M.~Salvatori, C.~A.~Scrucca and L.~Silvestrini,
  JHEP {\bf 0510}, 037 (2005)
  [arXiv:hep-ph/0503179].
G.~Cacciapaglia, C.~Csaki and S.~C.~Park,
  JHEP {\bf 0603}, 099 (2006)
  [arXiv:hep-ph/0510366]
 G.~Panico, M.~Serone and A.~Wulzer,
  Nucl.\ Phys.\ B {\bf 739}, 186 (2006)
  [arXiv:hep-ph/0510373].

\bibitem{6Drecenti}
G.~R.~Dvali, S.~Randjbar-Daemi, R.~Tabbash,
Phys.\ Rev.\ D {\bf 65} (2002) 064021
[hep-ph/0102307];
L.~J.~Hall, Y.~Nomura, D.~R.~Smith,
Nucl.\ Phys.\ B {\bf 639} (2002) 307
[hep-ph/0107331];
I.~Antoniadis, K.~Benakli, M.~Quiros,
New J.\ Phys.\  {\bf 3} (2001) 20
[hep-th/0108005];
C.~Csaki, C.~Grojean, H.~Murayama,
Phys.\ Rev.\ D {\bf 67} (2003) 085012
[hep-ph/0210133];
G.~von Gersdorff, N.~Irges and M.~Quiros,
  Phys.\ Lett.\ B {\bf 551} (2003) 351
  [arXiv:hep-ph/0210134].
C.~A.~Scrucca, M.~Serone, L.~Silvestrini and A.~Wulzer,
JHEP {\bf 0402}, 049 (2004)\newline
[arXiv:hep-th/0312267].
Y.~Hosotani, S.~Noda and K.~Takenaga,
Phys.\ Rev.\ D {\bf 69} (2004) 125014
[hep-ph/0403106];
 C.~Biggio and M.~Quiros,
  Nucl.\ Phys.\ B {\bf 703} (2004) 199
  [arXiv:hep-ph/0407348].

\bibitem{Randjbar-Daemi:1982hi}
  S.~Randjbar-Daemi, A.~Salam and J.~A.~Strathdee,
  Nucl.\ Phys.\ B {\bf 214} (1983) 491.


\bibitem{'tHooft:1979uj}
  G.~'t Hooft,
  Nucl.\ Phys.\ B {\bf 153}, 141 (1979).


\bibitem{Hosotani}
Y.~Hosotani,
Phys.\ Lett.\ B {\bf 126} (1983) 309;
Phys.\ Lett.\ B {\bf 129} (1983) 193;
Annals Phys.\  {\bf 190} (1989) 233.


\bibitem{ABGRS}
J.~Alfaro, A.~Broncano, M.~B.~Gavela, S.~Rigolin and M.~Salvatori,
  arXiv:hep-ph/0606070.


\bibitem{Ambjorn:1980sm}
  J.~Ambjorn and H.~Flyvbjerg,
  Phys.\ Lett.\ B {\bf 97}, 241 (1980).


\bibitem{diag_field_strenght}
  H.~Leutwyler,
  %
  Nucl.\ Phys.\ B {\bf 179}, 129 (1981).

 P.~van Baal,
  Commun.\ Math.\ Phys.\  {\bf 94}, 397 (1984).

\bibitem{thooft}
G.~'t Hooft,
  Commun.\ Math.\ Phys.\  {\bf 81} (1981) 267.

\bibitem{van_baal}
P.~van Baal,
Commun.\ Math.\ Phys.\  {\bf 85}, 529 (1982); 


\bibitem{Hernandez}
 A. F. Faedo, D. Hernandez, S. Rigolin and M. Salvatori, work in progress


\bibitem{Groeneveld:1980tt}
  J.~Groeneveld, J.~Jurkiewicz and C.~P.~Korthals Altes,
  Phys.\ Scripta {\bf 23}, 1022 (1981).

\bibitem{Gonzalez-Arroyo:1982hz}
  A.~Gonzalez-Arroyo and M.~Okawa,
  Phys.\ Rev.\  D {\bf 27}, 2397 (1983).

\bibitem{Gonzalez-Arroyo:1982ub}
  A.~Gonzalez-Arroyo and M.~Okawa,
  Phys.\ Lett.\  B {\bf 120}, 174 (1983).

\bibitem{vanBaal:1983eq}
  P.~van Baal,
  Commun.\ Math.\ Phys.\  {\bf 92}, 1 (1983).


\bibitem{Brihaye:1983yd}
  Y.~Brihaye, G.~Maiella and P.~Rossi,
  Nucl.\ Phys.\  B {\bf 222}, 309 (1983).

\bibitem{vanBaal:1985na}
  P.~van Baal and B.~van Geemen,
  J.\ Math.\ Phys.\  {\bf 27}, 455 (1986).

\bibitem{vanGeemen:1984wy}
  B.~van Geemen and P.~van Baal,
  Kon.\ Ned.\ Akad.\ Wetensch.\ Proc.\  B {\bf 89}, 39 (1986).



\bibitem{tony}
 A.Gonz\'alez Arroyo, 
 hep-th/9807108.

%




\end{thebibliography}
\end{document}